\documentclass[nonacm,acmsmall,natbib=false]{acmart}
\PassOptionsToPackage{hyphens}{url}\usepackage{hyperref}

\usepackage{xcolor}
\usepackage{xspace}
\usepackage{enumitem}
\usepackage{caption}

\usepackage{tikz}
\usepackage{pgfplots}
\pgfplotsset{compat=1.18}
\usepgfplotslibrary{fillbetween}
\usepgfplotslibrary{statistics}
\usepgfplotslibrary{colormaps}

\usepackage{pgfplotstable}
\usepackage{booktabs}
\usepackage{array}
\usepackage{adjustbox}

\usepackage{subcaption}

\makeatletter
\renewcommand\p@subfigure{} 
\makeatother

\newcommand{\myitem}[1]{\vspace*{0.02in}\noindent\textbf{#1}}

\def\showcomments{0}
\ifnum\showcomments=1

\newcommand{\kostas}[1]{{\footnotesize\color{purple}[kostas: #1]}}
\newcommand{\maria}[1]{{\footnotesize\color{blue}[maria: #1]}}

\else
\newcommand{\kostas}[1]{}
\newcommand{\maria}[1]{}

\fi

\newcommand{\remove}[1]{}

\newcommand{\ie}{\emph{i.e.,}\xspace}

\newcommand{\sys}{HOWLR\textsc{}\xspace}
\newcommand{\interception}{BGP-enabled impersonation\textsc{}\xspace}

\pgfplotsset{
    cycle list={
        {draw=blue!70!black},
        {draw=green!60!black},
        {draw=orange!80!black},
        {draw=red!70!black},
        {draw=purple!70!black}
    }
}
\pgfplotsset{
    every axis/.style={
        width=0.9\linewidth,
        height=0.5\linewidth,
        grid=both,
        grid style={line width=0.2pt, draw=gray!30},
        major grid style={line width=0.3pt, draw=gray!40},
        axis line style={black},
        tick style={black},
        tick label style={font=\small},
        label style={font=\small},
        title style={font=\small},
        legend style={font=\small, draw=none, fill=none},
        ymajorgrids=true,
        xmajorgrids=false,
        enlarge x limits=0.08,
        bar width=8pt,
    },
    compat=1.18,
}

\RequirePackage[
  datamodel=acmdatamodel,
  style=acmnumeric,
  ]{biblatex}
\begin{document}


\title{\sys: A Client-Driven Approach to BGP Hijack Detection}

\author{Constantine Doumanidis}
\author{Anya Kalogerakos}
\author{Maria Apostolaki}
\affiliation{%
  \institution{Princeton University}
  \city{Princeton}
  \state{New Jersey}
  \country{USA}
}

\begin{abstract}
BGP hijacking enables impersonation attacks in which adversaries divert traffic at the prefix level and serve malicious content to unsuspecting clients. Detecting such attacks has traditionally been the responsibility of network operators, leaving end hosts exposed for hours. We argue that end hosts can detect prefix-level impersonation independently, exploiting a fundamental asymmetry: a BGP hijack diverts traffic for an entire IP prefix, but impersonating every co-hosted service within that prefix is prohibitively difficult at scale, especially if each service is authenticated by a different Certificate Authority.
We propose HOWLR, a tool that operationalizes this insight by using co-hosted, TLS-authenticated services as witnesses: if a client can no longer authenticate them, it has evidence of an ongoing attack.
This work evaluates the feasibility of this method by quantifying the existence and diversity of witnesses in the wild. We show that HOWLR can protect 89\% of Tor relay prefixes, and 75\% of Bitcoin pool gateway prefixes.
\end{abstract}

\maketitle

\section{Introduction}


Because BGP does not authenticate route announcements by default, adversaries can divert traffic through their infrastructure — a BGP hijack. Despite numerous efforts to prevent them, such attacks remain common: in 2024 alone, BGPWatch reported 5,673 possible hijack events 
~\cite {bgpwatch}. 

Particularly dangerous are impersonation attacks, in which the attacker impersonates the legitimate destination rather than dropping traffic or returning it to the legitimate destination.
In impersonation attacks, the affected end hosts assume they are communicating with the legitimate destination, allowing adversaries to gain confidential information and/or inject malicious content.
For instance, a \interception attack targeting the Korean provider Kakao in 2022 allowed the attacker to serve malicious content, ultimately resulting in the theft of 2 million US dollars from users of KLAYswap~\cite{klayswap_bgp}.
Similarly, an attacker hijacked an Amazon prefix to obtain a valid TLS certificate for a Celer subdomain and redirect users of the Celer Bridge from the legitimate UI endpoint hosted on Amazon to attacker-controlled servers. As a result, users unknowingly approved malicious transactions, leading to the immediate theft of \$235,000 in user funds.

In a \interception attack, a client connected to the impersonated service has little ability to detect malicious behavior—especially when the communication lacks authentication or the attacker has already obtained a bogus certificate, as in the examples above.
Solutions for detecting BGP hijacks typically operate at the network level and are deployed by network operators. This is because most defenses rely on analyzing BGP advertisements, which requires access to \emph{diverse } BGP data feeds—resources that are generally unavailable to end-users, are slow and still suffer from blind spots~\cite{monitor_selection,evading_monitors}. 
Analyzing more accessible data, such as RouteViews~\cite{routeviews}, is not adequately diverse, is only updated every 15 minutes, and could create a bottleneck if used by all end-hosts.
Data-plane detections, which can be initiated by end hosts, such as sending traceroutes or pings, can be incomplete or subject to tampering~\cite{secure-traceroute}.

There is a clear and dangerous gap between the victims of \interception attacks (\ie clients unknowingly connected to bogus servers) and the entities capable of detecting them (\ie network providers). In this paper, we argue for a practical, end-host-based detection approach built on a simple but powerful observation: end hosts can detect \interception attacks by leveraging the mismatch between the broad impact of BGP hijacks and the limited scalability of successful impersonation. Because routers discard IP prefix announcements that exceed a certain maximum length (typically $/24$ for IPv4, and $/48$ for IPv6), an attacker cannot hijack a single IP address on which the target service is hosted. Instead, they must divert traffic for an entire $/24$ block, affecting at least $2^8$ host addresses. While diverting this many clients requires only a single BGP prefix announcement, impersonating them all is significantly harder. Concretely, the attacker would need to convince all their prospective connections that they are communicating with the legitimate destinations.
For destinations that host authenticated services, impersonation would require either compromising the service’s private key or deceiving a Certificate Authority (CA) into issuing a fraudulent certificate~\cite{birge2018bamboozling}. Although recent attacks and research have shown such compromises are possible, they do not scale to large numbers of servers. While an attacker might attempt to relay traffic for non-targeted services back to their legitimate destinations, doing so would require complex routing configurations and would constrain the attacker's location to highly connected ones who are unlikely to risk their reputation.

Leveraging these insights, we develop \sys, a detection mechanism that enables end hosts to detect whether a \interception attack is happening on a service they are connected to, independently of network-level monitoring. To achieve this, \sys connects to servers hosted at the same prefix as the service of interest and authenticates. We call such servers witnesses, and they are primarily TLS-enabled websites due to their reliability and prevalence. At runtime, a local daemon periodically validates these witnesses against a pre-established database; a simultaneous failure in witness authentication serves as a high-confidence alert of prefix-level interception, prompting the end host to take defensive action.

The feasibility of such an approach depends on the empirical density and diversity of witnesses, namely cryptographically authenticatable services, within any given destination prefix. We conduct a measurement study to characterize the cohabitation patterns of such witnesses across the broader routed IPv4 address space. We find that witnesses exist in a substantial fraction of service-hosting prefixes; crucially, all measured popular destination prefixes contain witnesses authenticated by at least two distinct Certificate Authorities, providing the necessary heterogeneity for robust detection.


\sys can effectively protect a broad range of security-sensitive applications. For instance, it protects 89\% of prefixes hosting Tor relays \textemdash a frequently targeted~\cite{tor-cmu, tor-kax17, tor-sslstrip} service that is vulnerable to BGP hijacks~\cite{sun2015raptor, vanbever2014anonymity}. Likewise, it protects 75\% of Bitcoin mining pool gateway prefixes, another high-value target for adversaries capable of routing attacks~\cite{apostolaki2017hijacking}.

\section{Background} \label{sec:sidechan}

We begin this section with a motivating example that illustrates the threat targeted by \sys. To make the motivation more engaging, we highlight a recent real-world attack rather than a hypothetical one. We then describe the threat model manifested in this attack and argue that it captures a broader class of practical threats. Finally, we discuss why existing detection mechanisms are inadequate—a fact underscored by the continued success of such attacks in the wild.


\begin{figure*}[t]
    \centering
    \begin{subfigure}[t]{0.49\linewidth}
        \centering
        \includegraphics[width=0.75\linewidth, trim={0.6cm 5.64cm 11.5cm 0},clip]{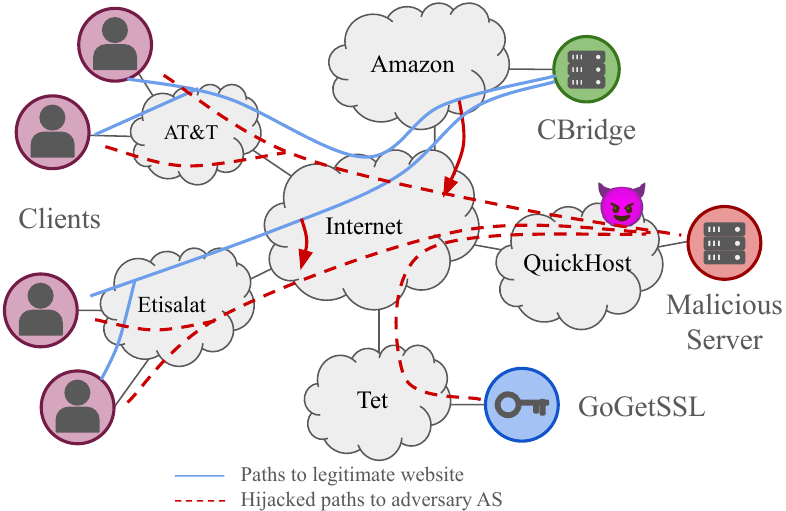}
        \caption{Real example of \interception attack~\cite{celerbridge_bgp}:
        The attacker, operating from QuickHost, launched a subprefix hijack to divert clients' requests to CBridge through her own server, impersonating the legitimate service. Before serving malicious content, she obtained a valid TLS certificate, making it appear as though clients were securely connected to the real site.}
        \label{fig:example-attack}
    \end{subfigure}
    \hfill
    \begin{subfigure}[t]{0.49\linewidth}
        \centering
        \includegraphics[width=1\linewidth, trim={0cm 6.8cm 10cm 0},clip]{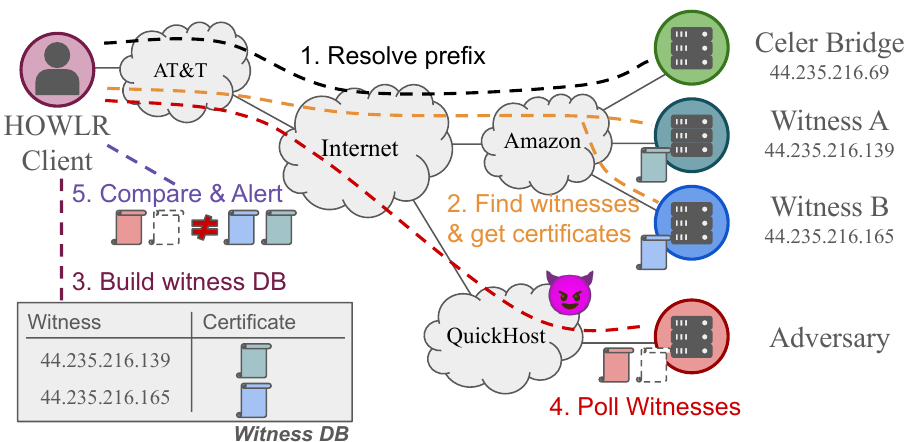}
        \caption{
        \sys resolves the IP address of the target service (Step 1), and enumerates its prefix to identify witnesses and retrieve their certificates (Step 2), which it stores in a local database (Step 3). \sys periodically polls witnesses for their certificates (Step 4), and compares retrieved and stored certificates (Step 5). \sys alerts the user if it detects discrepancies that indicate an interception attack.}
        \label{fig:example-howlr}
    \end{subfigure}

\end{figure*}

\subsection{Motivating Example}\label{sec:motivating-attack}
Figure~\ref{fig:example-attack} illustrates the infamous attack against the Celer bridge (CBridge) that resulted in 235,000 USD in assets being stolen~\cite{celerbridge_bgp}. CBridge is a service that facilitates quick and inexpensive cryptocurrency and assets transfers between different blockchains. The attacker targeted clients of CBridge, such as those displayed on the left side of Fig.~\ref{fig:example-attack}. These clients were originally using CBridge by visiting its front-end website, which was served by Amazon (AS 16509). 
During the attack, clients were redirected to an attacker-controlled server serving a malicious copy of the CBridge UI. 

To achieve this, the attacker first compromised a router belonging to QuickHost (AS 209243) and hijacked the Amazon prefix where CBridge's server was hosted (44.224.0.0/24). Having diverted traffic originally destined to CBridge to QuickHost, the attacker \emph{(i)} obtained a valid certificate from GoGetSSL, a trusted Certificate Authority hosted under Tet (AS 12578); and \emph{(ii)} served the clients whose traffic was diverted a malicious copy of the CBridge website that enabled the assets theft.

This example highlights two critical vulnerabilities in the Internet’s infrastructure in practice: \emph{(i)} BGP hijacks can succeed even against security-conscious operators, and \emph{(ii)} it is possible to acquire a bogus certificate for a legitimate website. While the exact mechanics are not central to our work, we provide additional context on how the attacker managed to divert Amazon traffic and acquire the bogus certificate.
First, the attacker was able to divert traffic to an Amazon prefix for approximately 3 hours, despite the fact that Amazon's prefixes are seemingly protected by RPKI. As explained by CertiK~\cite{celer-certik}, the attacker evaded detection by announcing a longer prefix (compared to the $/11$ originally advertised) with an Amazon AS as the origin, and QuickHost as the next hop.
Second, the attacker was able to obtain a fraudulent certificate by leveraging the ongoing BGP hijack. Specifically, she exploited the automated domain validation process to convince GoGetSSL—a trusted Certificate Authority—that she owned the CBridge domain. This allowed her to issue a valid TLS certificate for the impersonated service. This attack has been previously described in detail in ~\cite{birge2018bamboozling}.

\subsection{Threat Model}
The above attack generalizes to a broader threat model in which an attacker hijacks a prefix and impersonates a service. This model underpins several real-world incidents~\cite{myetherwallet_bgp, klayswap_bgp} and has been analyzed in the literature~\cite{birge2018bamboozling, gavrichenkov2015breaking}. We now break down the two steps of the model, namely BGP hijack and impersonation, and argue for the practicality of each.

\myitem{BGP hijack:} The attacker exercises control over a BGP-speaking router by either compromising the infrastructure of a network operator~\cite{dell_securewatch_bgp, orange2024bgp}, or by establishing her own Autonomous System (AS)~\cite{own-asn, nick-isp}. Although the latter option necessitates a modest financial investment - a conservative estimate places the annual operational costs for a small AS at approximately 2,500 USD - these expenditures can be recouped by the substantial revenue the attacker stands to generate as a result of her successful attack~\cite{klayswap_bgp}.
Using her access to a BGP-speaking router, the adversary issues malicious BGP announcements to hijack the IP prefixes of the services that she wishes to target. In particular, the adversary can perform more specific prefix or forged-origin subprefix hijacks that allow her to divert traffic destined for the victim prefixes to her own AS. 
The success of these attacks depends on the RPKI status of the victim prefixes. IP prefixes shorter than /24, that are not covered by a Route Origin Authorization (ROA) can be effortlessly targeted using a more specific prefix hijack. In this scenario, the attacker redirects all traffic to the victim prefix by announcing a longer prefix, exploiting the longest-prefix preference of BGP routers. IP prefixes covered by ROAs are also not immune to hijack attacks. ROAs are often observed to feature an overly permissive maxLength attribute, rendering many covered prefixes vulnerable to forged-origin subprefix hijack attacks~\cite{gilad2017maxlength}, allowing the adversary to again intercept all victim traffic. Even when a prefix is properly protected by RPKI in theory, it is not necessarily safe from hijack attacks in practice~\cite{cloudflare2024incident} since approximately 43\% of ASes currently do not enforce route origin validation~\cite{hlavacek2023keep}.
 The adversary can also attempt to completely evade RPKI validation by making an equally specific prefix announcement with the genuine origin, and itself prepended as a next hop, which will allow her to statistically divert 50\% of traffic destined for the victim prefix~\cite{securingApplications2021, goldbergSecuring}.

\myitem{Impersonation:} The attacker’s goal is to convince clients connected to the targeted service or host within the hijacked prefix that they are communicating with the legitimate destination. For services that do not employ some form of authentication, such as DNS servers without DNSSEC or web servers without TLS certificates, this becomes trivial. The attacker simply responds to connected clients with malicious content. For services that employ some form of server authentication, such as websites, the attacker can obtain access to the private key to the certificate served by the victim service, or convince a legitimate Certificate Authority (CA) to issue a new one for her~\cite{birge2018bamboozling}. While recent incidents have shown that fooling a CA into issuing a fraudulent certificate is possible, doing so for all co-hosted services within a prefix is not practical: it requires independently repeating the attack for each domain and even each relevant CA. Moreover, CAs that do not rely on automated domain validation are altogether immune to this technique, making full-prefix impersonation infeasible in practice. We therefore consider this out of scope for our threat model.

\subsection{Traditional Hijack Detection Limitations} 
Traditional BGP hijack detection relies on network operators, is often slow, and is agnostic to the criticality of the targeted services. Many approaches require access to diverse BGP data streams~\cite{mai2008detecting, li2005, huang2007diagnosing}, multiple vantage points~\cite{argus, hu2007accurate}, and collaboration with other ASes~\cite{haeberlen2009netreview} \textemdash resources typically inaccessible to end users. Detection is also slow, since many methods rely on wide temporal windows~\cite{mai2008detecting, al2015detecting, prakash2009bgp} or on aggregated BGP data sources like RouteViews~\cite{routeviews}, which publishes BGP updates only every 15 minutes. Even in the case of methods that can be adapted to run in real-time using live BGP data feeds, researchers expect increased false positives~\cite{shapira2022ap2vec}. Finally, these methods are oblivious to the specific applications being targeted, making the timely detection, user alert, and incident response process inefficient.

Independent of the detection method, mitigation remains a slow and manual process that leaves users exposed for prolonged periods of time. This is because removing malicious routes from BGP routing tables often requires manual coordination between network operators, such as contacting affected ASes and installing filters. Consequently, response times often lag for hours, as seen in recent high-profile incidents~\cite{celerbridge_bgp, orange2024bgp}. During this time, users of the targeted services remain unaware and vulnerable. Incidents such as the one in \S\ref{sec:motivating-attack} highlight the urgent need to make timely hijack detection accessible to end hosts.

\section{Overview}
\label{sec:overview}

In this section, we introduce \sys, a lightweight detection mechanism that enables a client to identify when its connection to a target service is being impersonated. \sys relies on the insight that during an interception attack, services hosted in the same IP prefix as the target service cannot be authenticated. 
By detecting this failure, the client can take timely action to mitigate the risk. For instance, in a web browser, this could mean preventing a spoofed malicious website from loading; in a P2P network it could mean removing the impersonated peer from the active peer set.
\sys requires only that the client be able to initiate connections to other hosts within the target service prefix.

\sys operates in two steps: \textit{witness discovery}, and \textit{active monitoring}. At initialization, \sys takes as input the target service IP prefix-specifically, the longest prefix that BGP routers will propagate (/24 for IPv4 or /48 for IPv6). \sys then enumerates the IP addresses in the prefix and attempts to connect to the corresponding hosts. For each responsive host, it scans it for open ports, and then attempts to perform a TLS handshake on each open port to retrieve any offered certificates. \sys validates the certificates it retrieves, and adds valid ones and their corresponding hosts to its witness database for the target prefix. 

Once witness discovery is complete, \sys actively monitors witnesses by periodically, or on-demand, performing TLS handshakes with them and comparing the retrieved certificates with the ones that in the witness database. If \sys cannot authenticate the witnesses ---e.g., due to certificate mismatches or failed connections--- \sys raises an alert, triggering a response according to the client's configuration.


\myitem{Example application:} We retroactively apply \sys to the CBridge incident discussed in Section~\ref{sec:motivating-attack} and show how it could have protected clients. Suppose that the client under AT\&T in Figure~\ref{fig:example-howlr} deployed \sys when it first accessed the bridge prior to the attack. 
In \textbf{Step 1} \sys observes the domain names of CBridge that the client loads, and resolves them to their IP addresses. 
In \textbf{Step 2} \sys enumerates the IP addresses in Amazon's 44.235.216.0/24 prefix, where the bridge is hosted, and looks for witnesses. 
Among others, it identifies witnesses A and B, the web servers of PIXELL AI, a video enhancement SaaS, and AdRoll, an advertising platform, and obtains their certificates, issued by Amazon and GlobalSign.
While we lack historical certificate data from the time of the attack, scans today suggest that multiple witnesses would have been available.
In \textbf{Step 3}, \sys stores the IP addresses, certificates, and identification timestamps of the witnesses in its database. 
Now assume that the adversary launches her hijack by announcing 44.235.216.0/24 as described in Section~\ref{sec:motivating-attack}. In \textbf{Step 4}, when \sys polls witnesses, its traffic will be redirected to QuickHost. The attacker can either drop \sys's traffic or attempt to serve bogus certificates obtained by fooling a CA into issuing new ones~\cite{birge2018bamboozling}. In \textbf{Step 5}, \sys detects the mismatch between the adversary's certificate set and the one in its database and raises an alert, preventing the user from accessing the malicious bridge website, which would have led to the theft of their assets.






\section{Measuring Witnesses in the wild}
\label{sec:measurements}
\begin{figure*}[t]
    \centering
    \begin{subfigure}[t]{0.49\linewidth}
        \centering
        \begin{tikzpicture}
            \begin{axis}[
                ybar,
                enlarge x limits=0.25,
                legend pos=north west,
                legend style={at={(0.45,1.02)},anchor=north,draw=none, fill=none, legend columns=-1,font=\footnotesize},
                ylabel={Count},
                xlabel={Application},
                symbolic x coords={BTC Nodes, Pool Gateways, Tor Relays},
                xtick=data,
                x tick label style={font=\footnotesize},
                bar width=10pt,
                width=1\linewidth,
                height=0.5\linewidth,
                xlabel={Application},
                ylabel={IP Prefixes},
                ymin=0,
                xlabel style={yshift=4pt,},
            ]
            
            \addplot[fill=red!80!black, draw=black] coordinates {
                (BTC Nodes,17) 
                (Pool Gateways,11) 
                (Tor Relays,3)
            };
            
            \addplot[fill=blue!80!black, draw=black] coordinates {
                (BTC Nodes,2) 
                (Pool Gateways,5) 
                (Tor Relays,13)
            };
            
            \addplot[fill=green!60!black, draw=black] coordinates {
                (BTC Nodes,31) 
                (Pool Gateways,28) 
                (Tor Relays,35)
            };
            
            \legend{Unprotected, Light, Strong}
            
            \end{axis}
        \end{tikzpicture}
        \caption{\sys can protect the prefixes of 65\% of Bitcoin nodes, 75\% of pool gateways and 96\% of Tor relays.}
        \label{fig:applications}
    \end{subfigure}
    \hfill
    \begin{subfigure}[t]{0.49\linewidth}
        \centering
        \begin{tikzpicture}
            \begin{axis}[
                width=\linewidth,
                height=0.5\linewidth,
                xlabel={Number of Witnesses},
                ylabel={Prefixes CDF},
                grid=both,
                xmin=0,
                ymin=0,
                ymax=1,
                ymax=1.05,
                legend pos=south east,
                legend style={draw=none, fill=none},
                tick label style={font=\small},
                label style={font=\small},
            ]
            
            \addplot+[
                line width=1.2pt,
            ]table{figures/cdf_random.dat};
            \addlegendentry{Random Prefixes}
            
            \addplot+[
                line width=1.2pt,
                dashed,
            ]table{figures/cdf_popular.dat};
            \addlegendentry{Popular Services}
            
            \end{axis}
        \end{tikzpicture}
        \caption{Prefixes that host popular services are more likely to have witnesses, compared to random prefixes.}
        \label{fig:witnesses-cdf}
    \end{subfigure}
\end{figure*}
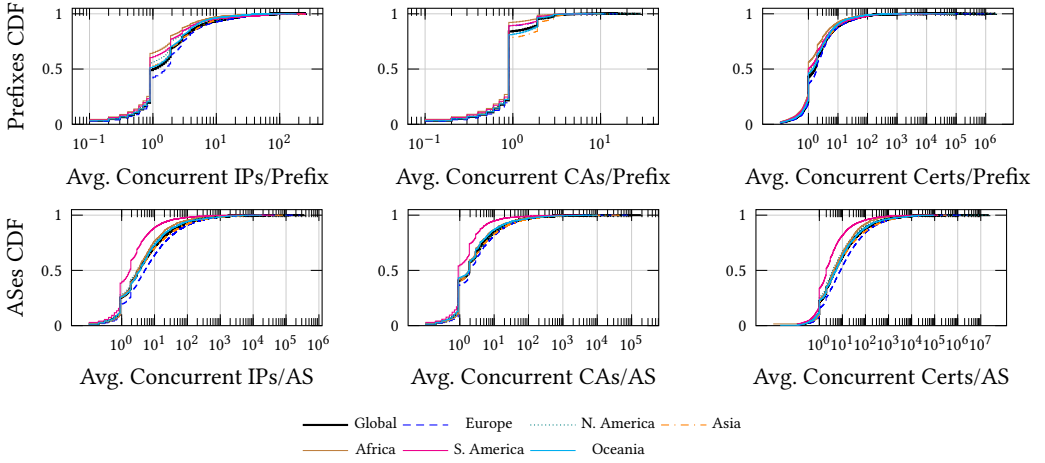
\begin{figure*}[t]
    \pgfplotsset{
        every axis/.append style={
            width=1.1\linewidth,
            height=0.65\linewidth,
            grid=major,
            ymin=0, ymax=1.05,
            label style={font=\small},
            tick label style={font=\tiny},
            legend style={font=\tiny, at={(0.5,-0.25)}, anchor=north, legend columns=4},
        },
        global_line/.style={black, thick},
        eu_line/.style={blue, densely dashed},
        na_line/.style={teal, densely dotted},
        as_line/.style={orange, dashdotted},
        af_line/.style={brown},
        sa_line/.style={magenta},
        oc_line/.style={cyan}
    }
    \centering
    \begin{subfigure}[t]{0.32\linewidth}
        \centering
        \begin{tikzpicture}
        \begin{axis}[
            width=1.1\linewidth,
            height=0.7\linewidth,
            xlabel={Avg. Concurrent IPs/Prefix},
            ylabel={Prefixes CDF},
            grid=major,
            xmode=log,
            ymin=0, ymax=1.05,
            legend to name=regional-legend,
        ]
            
            \addplot [global_line, const plot] table [x=value, y=cdf, col sep=comma] {data/coverage/all/witness_coverage_ips_cdf.csv};
            \addlegendentry{Global}
            
            \addplot [eu_line, const plot] table [x=value, y=cdf, col sep=comma] {data/coverage/europe/witness_coverage_Europe_ips_cdf.csv};
            \addlegendentry{Europe}
            
            \addplot [na_line, const plot] table [x=value, y=cdf, col sep=comma] {data/coverage/north_america/witness_coverage_North_America_ips_cdf.csv};
            \addlegendentry{N. America}
            
            \addplot [as_line, const plot] table [x=value, y=cdf, col sep=comma] {data/coverage/asia/witness_coverage_Asia_ips_cdf.csv};
            \addlegendentry{Asia}
            
            \addplot [af_line, const plot] table [x=value, y=cdf, col sep=comma] {data/coverage/africa/witness_coverage_Africa_ips_cdf.csv};
            \addlegendentry{Africa}
            
            \addplot [sa_line, const plot] table [x=value, y=cdf, col sep=comma] {data/coverage/south_america/witness_coverage_South_America_ips_cdf.csv};
            \addlegendentry{S. America}
            
            \addplot [oc_line, const plot] table [x=value, y=cdf, col sep=comma] {data/coverage/oceania/witness_coverage_Oceania_ips_cdf.csv};
            \addlegendentry{Oceania}
        \end{axis}
        \end{tikzpicture}
    \end{subfigure}
    \hfill
    \begin{subfigure}[t]{0.32\linewidth}
        \centering
        \begin{tikzpicture}
        \begin{axis}[
            width=1.1\linewidth,
            height=0.7\linewidth,
            xlabel={Avg. Concurrent CAs/Prefix},
            grid=major,
            xmode=log,
            ymin=0, ymax=1.05,
        ]

            \addplot [global_line, const plot] table [x=value, y=cdf, col sep=comma] {data/coverage/all/witness_coverage_cas_cdf.csv};
            
            \addplot [eu_line, const plot] table [x=value, y=cdf, col sep=comma] {data/coverage/europe/witness_coverage_Europe_cas_cdf.csv};
            
            \addplot [na_line, const plot] table [x=value, y=cdf, col sep=comma] {data/coverage/north_america/witness_coverage_North_America_cas_cdf.csv};
            
            \addplot [as_line, const plot] table [x=value, y=cdf, col sep=comma] {data/coverage/asia/witness_coverage_Asia_cas_cdf.csv};
            
            \addplot [af_line, const plot] table [x=value, y=cdf, col sep=comma] {data/coverage/africa/witness_coverage_Africa_cas_cdf.csv};
            
            \addplot [sa_line, const plot] table [x=value, y=cdf, col sep=comma] {data/coverage/south_america/witness_coverage_South_America_cas_cdf.csv};
            
            \addplot [oc_line, const plot] table [x=value, y=cdf, col sep=comma] {data/coverage/oceania/witness_coverage_Oceania_cas_cdf.csv};
        \end{axis}
        \end{tikzpicture}
    \end{subfigure}
    \hfill
    \begin{subfigure}[t]{0.32\linewidth}
        \centering
        \begin{tikzpicture}
        \begin{axis}[
            width=1.1\linewidth,
            height=0.7\linewidth,
            xlabel={Avg. Concurrent Certs/Prefix},
            grid=major,
            xmode=log,
            ymin=0, ymax=1.05,
            xtick={1, 10, 100, 1000, 10000, 100000, 1000000},
            x tick label style={font=\tiny},
        ]

            \addplot [global_line, const plot] table [x=value, y=cdf, col sep=comma] {data/coverage/all/witness_coverage_fps_cdf.csv};
            
            \addplot [eu_line, const plot] table [x=value, y=cdf, col sep=comma] {data/coverage/europe/witness_coverage_Europe_fps_cdf.csv};
            
            \addplot [na_line, const plot] table [x=value, y=cdf, col sep=comma] {data/coverage/north_america/witness_coverage_North_America_fps_cdf.csv};
            
            \addplot [as_line, const plot] table [x=value, y=cdf, col sep=comma] {data/coverage/asia/witness_coverage_Asia_fps_cdf.csv};
            
            \addplot [af_line, const plot] table [x=value, y=cdf, col sep=comma] {data/coverage/africa/witness_coverage_Africa_fps_cdf.csv};
            
            \addplot [sa_line, const plot] table [x=value, y=cdf, col sep=comma] {data/coverage/south_america/witness_coverage_South_America_fps_cdf.csv};
            
            \addplot [oc_line, const plot] table [x=value, y=cdf, col sep=comma] {data/coverage/oceania/witness_coverage_Oceania_fps_cdf.csv};
        \end{axis}
        \end{tikzpicture}
    \end{subfigure}
    \hfill
    \begin{subfigure}[t]{0.32\linewidth}
        \centering
        \begin{tikzpicture}
        \begin{axis}[
            width=1.1\linewidth,
            height=0.7\linewidth,
            xlabel={Avg. Concurrent IPs/AS},
            ylabel={ASes CDF},
            grid=major,
            xmode=log,
            ymin=0, ymax=1.05,
            xtick={1, 10, 100, 1000, 10000, 100000, 1000000},
        ]
            
            \addplot [global_line, const plot] table [x=value, y=cdf, col sep=comma] {data/coverage/asn/witness_coverage_asn_ips_cdf.csv};
            
            \addplot [eu_line, const plot] table [x=value, y=cdf, col sep=comma] {data/coverage/europe_asn/witness_coverage_Europe_asn_ips_cdf.csv};
            
            \addplot [na_line, const plot] table [x=value, y=cdf, col sep=comma] {data/coverage/north_america_asn/witness_coverage_North_America_asn_ips_cdf.csv};
            
            \addplot [as_line, const plot] table [x=value, y=cdf, col sep=comma] {data/coverage/asia_asn/witness_coverage_Asia_asn_ips_cdf.csv};
            
            \addplot [af_line, const plot] table [x=value, y=cdf, col sep=comma] {data/coverage/africa_asn/witness_coverage_Africa_asn_ips_cdf.csv};
            
            \addplot [sa_line, const plot] table [x=value, y=cdf, col sep=comma] {data/coverage/south_america_asn/witness_coverage_South_America_asn_ips_cdf.csv};
            
            \addplot [oc_line, const plot] table [x=value, y=cdf, col sep=comma] {data/coverage/oceania_asn/witness_coverage_Oceania_asn_ips_cdf.csv};
        \end{axis}
        \end{tikzpicture}
    \end{subfigure}
    \hfill
    \begin{subfigure}[t]{0.32\linewidth}
        \centering
        \begin{tikzpicture}
        \begin{axis}[
            width=1.1\linewidth,
            height=0.7\linewidth,
            xlabel={Avg. Concurrent CAs/AS},
            grid=major,
            xmode=log,
            ymin=0, ymax=1.05,
            xtick={1, 10, 100, 1000, 10000, 100000, 1000000},
        ]
            
            \addplot [global_line, const plot] table [x=value, y=cdf, col sep=comma] {data/coverage/asn/witness_coverage_asn_cas_cdf.csv};
            
            \addplot [eu_line, const plot] table [x=value, y=cdf, col sep=comma] {data/coverage/europe_asn/witness_coverage_Europe_asn_cas_cdf.csv};
            
            \addplot [na_line, const plot] table [x=value, y=cdf, col sep=comma] {data/coverage/north_america_asn/witness_coverage_North_America_asn_cas_cdf.csv};
            
            \addplot [as_line, const plot] table [x=value, y=cdf, col sep=comma] {data/coverage/asia_asn/witness_coverage_Asia_asn_cas_cdf.csv};
            
            \addplot [af_line, const plot] table [x=value, y=cdf, col sep=comma] {data/coverage/africa_asn/witness_coverage_Africa_asn_cas_cdf.csv};
            
            \addplot [sa_line, const plot] table [x=value, y=cdf, col sep=comma] {data/coverage/south_america_asn/witness_coverage_South_America_asn_cas_cdf.csv};
            
            \addplot [oc_line, const plot] table [x=value, y=cdf, col sep=comma] {data/coverage/oceania_asn/witness_coverage_Oceania_asn_cas_cdf.csv};
        \end{axis}
        \end{tikzpicture}
    \end{subfigure}
    \hfill
    \begin{subfigure}[t]{0.32\linewidth}
        \centering
        \begin{tikzpicture}
        \begin{axis}[
            width=1.1\linewidth,
            height=0.7\linewidth,
            xlabel={Avg. Concurrent Certs/AS},
            grid=major,
            xmode=log,
            ymin=0, ymax=1.05,
            xtick={1, 10, 100, 1000, 10000, 100000, 1000000, 10000000},
        ]
            
            \addplot [global_line, const plot] table [x=value, y=cdf, col sep=comma] {data/coverage/asn/witness_coverage_asn_fps_cdf.csv};
            
            \addplot [eu_line, const plot] table [x=value, y=cdf, col sep=comma] {data/coverage/europe_asn/witness_coverage_Europe_asn_fps_cdf.csv};
            
            \addplot [na_line, const plot] table [x=value, y=cdf, col sep=comma] {data/coverage/north_america_asn/witness_coverage_North_America_asn_fps_cdf.csv};
            
            \addplot [as_line, const plot] table [x=value, y=cdf, col sep=comma] {data/coverage/asia_asn/witness_coverage_Asia_asn_fps_cdf.csv};
            
            \addplot [af_line, const plot] table [x=value, y=cdf, col sep=comma] {data/coverage/africa_asn/witness_coverage_Africa_asn_fps_cdf.csv};
            
            \addplot [sa_line, const plot] table [x=value, y=cdf, col sep=comma] {data/coverage/south_america_asn/witness_coverage_South_America_asn_fps_cdf.csv};
            
            \addplot [oc_line, const plot] table [x=value, y=cdf, col sep=comma] {data/coverage/oceania_asn/witness_coverage_Oceania_asn_fps_cdf.csv};
        \end{axis}
        \end{tikzpicture}
    \end{subfigure}
    \hfill
    \vspace{0.2em}
    \ref{regional-legend} 
    \caption{Distribution of IPv4 prefixes / ASes over average concurrent witness IPs/CAs/Certificates for witness-hosting prefixes / ASes between April 15 - May 15, 2026.}
    \label{fig:witness_coverage_cdfs}
\end{figure*}

\sys introduces a client-driven approach to detecting BGP-enabled impersonation by validating the integrity of an entire IP prefix, rather than individual hosts. This is made possible through \emph{witnesses}—services that cohabitate a prefix and serve as indirect signals of authenticity. In this section, we outline the requirements for such witness-based detection to succeed, and demonstrate through measurements that these requirements are met across real-world prefixes.

\subsection*{Requirements for Witness-Based Validation}
For \sys to reliably detect prefix-level impersonation attacks, clients must be able to identify a set of witnesses that satisfy three core properties:

\myitem{Prefix-locality:} The witness must reside within the same /24 prefix as the target service. Since BGP hijacks are typically carried out at the prefix level, prefix-local witnesses will be co-affected by a hijack, enabling detection through shared inconsistencies.

\myitem{Authenticatability:} The witness must expose a cryptographic identity—such as a TLS certificate or SSH key—that \sys can verify. Public key infrastructure (PKI), and in particular CA-signed TLS certificates, are attractive in this context: they are widely deployed, easy to validate through standard handshakes, and difficult to obtain fraudulently at scale.

\myitem{Operational Independence:} To ensure resilience, the witness set should ideally span multiple administrative domains and CAs. This redundancy increases both availability by avoiding coordinated outages (e.g., maintenance, power loss, DoS attacks) and robustness against compromise (e.g., a single CA being tricked into issuing fraudulent certificates~\cite{birge2018bamboozling}).

\subsection*{Measuring Witnesses Availability}
\begin{figure*}[t]
\centering
    \begin{subfigure}[t]{0.32\linewidth}
    \centering
    \begin{tikzpicture}[baseline=(current bounding box.north)]
    \begin{axis}[
        ybar,
        xlabel={Certificate CAs in /24 Prefix},
        ylabel={IP Prefixes},
        xtick=data,
        symbolic x coords={1,2,3,4,5,6,7,8,9,10,11},
        ymin=0,
        bar width=6pt,
        width=\linewidth,
        height=0.65\linewidth,
        x tick label style={font=\footnotesize},
    ]
    \addplot coordinates {
        (1,30) (2,14) (3,4) (4,3) (5,1)
        (6,5) (7,3) (8,4) (9,3) (10,0) (11,3)
    };
    \end{axis}
    \end{tikzpicture}
    \caption{57\% of sampled /24 prefixes with witnesses have certificates issued by 2+ unique CAs.}
    \label{fig:ca-diversity}
    \end{subfigure}
    \hfill
    \begin{subfigure}[t]{0.32\linewidth}
    \centering
    \begin{tikzpicture}[baseline=(current bounding box.north)]
    \begin{axis}[
        ybar,
        xlabel={Certificates per IP},
        ylabel={IP Addrs.},
        xtick=data,
        symbolic x coords={1,2,3,4,5,6,7,8,9,10,11,12,66},
        ymin=0,
        bar width=5pt,
        width=\linewidth,
        height=0.65\linewidth,
        x tick label style={rotate=0, font=\tiny},
    ]
    \addplot coordinates {
        (1,2522) (2,369) (3,30) (4,6) (5,3)
        (6,76) (7,17) (8,8) (9,16) (10,1)
        (11,1) (12,10) (66,1)
    };
    \end{axis}
    \end{tikzpicture}
    \caption{More than 82\% of hosts with certificates only serve a single certificate.}
    \label{fig:certificates-number}
    \end{subfigure}
    \hfill
    \begin{subfigure}[t]{0.32\linewidth}
    \centering
    \begin{tikzpicture}[baseline=(current bounding box.north)]
    \begin{axis}[
        ybar,
        xlabel={Port Number},
        ylabel={Certificates},
        xtick=data,
        symbolic x coords={443,2083,2087,2096,8443,2053,995,993,465,2078,2080,8883,5228,5030,5031},
        ymin=0,
        bar width=4pt,
        width=\linewidth,
        height=0.65\linewidth,
        x tick label style={rotate=90, anchor=east, font=\scriptsize},
        xlabel style={yshift=5pt},
    ]
    \addplot coordinates {
        (443,3354) (2083,119) (2087,119) (2096,118) (8443,103)
        (2053,95) (995,29) (993,28) (465,26) (2078,25)
        (2080,24) (8883,17) (5228,13) (5030,10) (5031,10)
    };
    \end{axis}
    \end{tikzpicture}
    \vspace{-0.15cm}
    \caption{82\% of certificates are served on port 443. Other ports contributed minimally.}
    \label{fig:certificates-ports}
    \end{subfigure}
\end{figure*}
We now examine whether real-world prefixes satisfy these requirements. Specifically, we evaluate the availability of prefix-local, authenticatable, and operationally independent witnesses across the IPv4 prefix space.

\myitem{Methodology:} A comprehensive investigation of witness availability requires enumerating over all IP addresses, establishing reachability to the corresponding hosts, iterating over all ports, and attempting TLS handshakes to retrieve certificates and verify their validity. Tools like ZMap~\cite{durumeric2013zmap} can perform internet-wide scans, however, doing so at our desired granularity would be impractical given the time and resource requirements. Furthermore, such scans require special considerations as they can overwhelm the scanned networks and hosts, and can trigger Denial-of-Service (DoS) defenses that undermine the fidelity of the results. 

We address these challenges using a two pronged approach: (i) We perform a small scale, but comprehensive, low-rate port scan and certificate retrieval over a sample of 100 /24 prefixes: 50 that host popular online services~\cite{explodingtopics,wikipedia-most-visited}, and 50 that are randomly drawn~\cite{ipvoid} and expose at least one public service. (ii) We process 5 months of newly issued certificates that were submitted to the Let's Encrypt "Willow" and Cloudflare "Nimbus" Certificate Transparency (CT) logs. We validate these certificates and identify the hosts that serve them via DNS resolution. Our small scale scan gives us witness availability insights at a fine grain, while processing CT logs allows us to answer questions on broad scale witnesses availability.

\myitem{Prefix-local and authenticatable witnesses:}
Our fine-grain scan indicates high witness availability for services hosted on common infrastructure, and more limited coverage of randomly selected prefixes. As shown in Figure~\ref{fig:witnesses-cdf}, all popular-service prefixes have at least two TLS-authenticated witnesses, and more than half contain 35 or more. By contrast, only 40\% of random prefixes feature any valid witnesses. This gap suggests that service popularity correlates with co-residency on richly instrumented infrastructure—an encouraging trend for protecting high-value targets.

Our CT log scans yield further insights about the distribution of witnesses across witness-hosting prefixes and ASes: 30\% of /24 witness prefixes are home to at least 3 certificate-hosting IPs at any given time, and 58\% of witness prefixes host more than a single certificate, as shown in Figure \ref{fig:witness_coverage_cdfs}. On the AS level, 50\% of witness ASes host at least 3 witnesses, with 49\% of them hosting at least 5 different certificates. However, trends on both prefix and AS levels do not generalize across regions. As can be seen in Figure \ref{fig:witness_coverage_cdfs}, witness-hosting prefixes and ASes in Europe tend to host more certificates compared to witness prefixes and ASes in Africa and South America. For example, while 50\% of European witness ASes host 8 or more certificates, only 10\% of witness ASes in Africa do so.

\myitem{Overall coverage:} Our study of CT log certificates indicates a limited spread of witnesses across the internet. In particular, witnesses from our CT log scan span just 17.4\% of the routable IPv4 space. This can be attributed to internet service centralization, and the limitations of the CT log measurement method. Services on the internet that serve TLS certificates tend to be hosted in large hosting and infrastructure providers, rather than be distrubuted across the broader network, which concentrates witnesses in a smaller subset of the IP prefix space. Furthermore, our methodology is limited by its source of data: while we process certificates from two of the largest CT logs, not all issued certificates are included in them. Furthermore, certificates are resolved to their prefixes via DNS resolution which can bias results through mechanisms such as GeoDNS, Global Server Load Balancing (GSLB), and the EDNS Client Subnet (ECS) extension. Regardless, our CT log study provides an informed lower bound on internet-wide witness coverage.

While counterintuitive, our CT log observations allow us to confirm the efficacy of our approach. In particular, we find that HOWLR can detect 83\% of real world prefix hijacks that took place in May 2026. We derive this by obtaining a dataset of the 1902 prefixes observed in high-confidence BGP hijack incidents through Cloudflare Radar~\cite{cloudflare-radar}.

\myitem{Authentication granularity:}
We also use our fine-grain scans to examine whether individual hosts offer multiple certificates, which could increase redundancy within a prefix. As Figure~\ref{fig:certificates-number} shows, 82\% of certificate-bearing hosts serve only a single certificate — implying that host diversity, not intra-host redundancy, is key. Finally, we wish to understand what services are better suited to act as witnesses. We find that 82\% of valid certificates are served on port 443 (Fig.~\ref{fig:certificates-ports}), typically used for HTTPS traffic.

Taken together, these results suggest that the requirements of prefix-locality and authenticatability are broadly satisfied in the wild—especially for service-rich prefixes.

\myitem{Operational independence via CA diversity:}
CA hosting provides an added layer of protection: attackers must deceive or compromise multiple CAs to convincingly impersonate an entire witness set.
To quantify the existence of independent witnesses, we analyze the diversity of CAs observed within each prefix. Our fine-grain scans show that 57\% of prefixes in our sample contain certificates issued by at least two distinct CAs (Figure~\ref{fig:ca-diversity}). At the same time, our CT log scans indicate that just 4\% of witness prefixes host certificates from two or more CAs. These results indicate that CA diversity is mostly observed in environments that host popular services. 

\section{\sys Design}
\pgfplotscreateplotcyclelist{my_protection_colors}{
    {fill=blue!70!black, draw=black},   
    {fill=green!60!black, draw=black},  
    {fill=orange!80!black, draw=black}, 
    {fill=red!70!black, draw=black},
    {fill=purple!70!black, draw=black} 
}
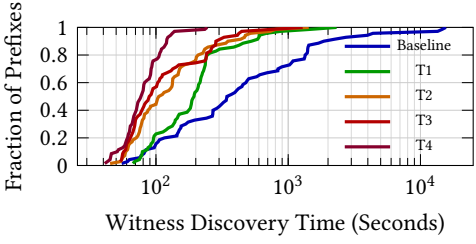
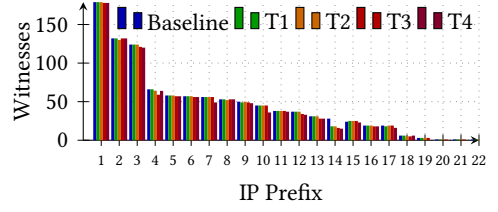
\begin{figure*}

    \centering
    \begin{subfigure}[t]{0.49\linewidth}
        \centering
        \begin{tikzpicture}
        \begin{axis}[
            width=1\linewidth,
            height=0.5\linewidth,
            xlabel={Witness Discovery Time (Seconds)},
            ylabel={Fraction of Prefixes},
            xmode=log,
            ymin=0, ymax=1,
            grid=both,
            legend pos=south east,
            legend style={draw=none, fill=none, font=\tiny, anchor=west,at={(0.65, 0.5)}},
            tick label style={font=\small},
            label style={font=\small},
        ]
        
        \addplot+[line width=1.2pt] table {figures/runtime_Baseline.dat};
        \addlegendentry{Baseline}
        
        \addplot+[line width=1.2pt] table {figures/runtime_T1.dat};
        \addlegendentry{T1}
        
        \addplot+[line width=1.2pt] table {figures/runtime_T2.dat};
        \addlegendentry{T2}
        
        \addplot+[line width=1.2pt] table {figures/runtime_T3.dat};
        \addlegendentry{T3}
        
        \addplot+[line width=1.2pt] table {figures/runtime_T4.dat};
        \addlegendentry{T4}
        
        \end{axis}
        \end{tikzpicture}
        \caption{Witness discovery time for variants: T1-stop host scan on certificate found, T2-also port 443 first, T3-also skip port 80 if no cert. on 443, T4-only scan 443.}
        \label{fig:runtime-optimizations}
    \end{subfigure}
    \hfill
    \begin{subfigure}[t]{0.49\linewidth}
        \centering
        \begin{tikzpicture}
        \begin{axis}[
            width=1\linewidth, height=0.5\linewidth,
            ybar=0pt,               
            bar width=1.2pt,         
            cycle list name=my_protection_colors,
            xlabel={IP Prefix},
            ylabel={Witnesses},
            ylabel shift=-5pt,
            xmin=-1,
            grid=major,
            major grid style={dotted, gray},
            axis lines=left,
            xtick=data,
            xticklabel={\pgfmathparse{int(\tick+1)}\pgfmathresult},
            x tick label style={font=\tiny},
            ymin=0,
            legend style={
                at={(1.02, 0.85)},   
                anchor=east,        
                legend columns=-1,   
                cells={anchor=west}, 
                font=\small,
                draw=none          
            },
            ]
        
        \addplot+[draw=none] 
            table [x expr=\coordindex, y=Baseline, col sep=comma] {figures/cert_counts_data.csv};
        
        \addplot+[draw=none] 
            table [x expr=\coordindex, y=T1, col sep=comma] {figures/cert_counts_data.csv};
        
        \addplot+[draw=none] 
            table [x expr=\coordindex, y=T2, col sep=comma] {figures/cert_counts_data.csv};
        
        \addplot+[draw=none] 
            table [x expr=\coordindex, y=T3, col sep=comma] {figures/cert_counts_data.csv};
        
        \addplot+[draw=none] 
            table [x expr=\coordindex, y=T4, col sep=comma] {figures/cert_counts_data.csv};
        
        \legend{Baseline, T1, T2, T3, T4}
        
        \end{axis}
        \end{tikzpicture}
        \caption{\sys variants miss a minimal number of witnesses compared to the baseline.}
        \label{fig:optimizations-drops}
    \end{subfigure}
\end{figure*}
Having shown that real-world prefixes offer sufficient witness coverage, we turn to validating whether \sys can operate efficiently in practice. We implement a lightweight client-side daemon that performs witness discovery and periodic validation using public scan data.

\myitem{Minimizing overhead:}
To reduce scanning cost, \sys integrates with public host indexing platforms (e.g., Censys~\cite{censys15} or CT logs~\cite{cloudflare_nimbus_ctlogs, letsencrypt_ctlogs}) to pre-filter for responsive hosts and open ports to perform TLS handshakes, minimizing unnecessary probing.

\myitem{Improving efficiency:}
To improve witness discovery efficiency, we incorporate two practical tweaks into our implementation, each informed directly by our earlier measurement findings (\S\ref{sec:measurements}). First, we adopt an early-stopping strategy: once a host yields a valid certificate, \sys halts further port probing for that host. This is motivated by our observation that 82\% of hosts with certificates serve only one (Figure~\ref{fig:certificates-number}), meaning additional probes are unlikely to uncover further useful authentication signals. Second, we prioritize port 443 during scanning, as it accounts for the vast majority (82\%) of discovered certificates (Figure~\ref{fig:certificates-ports}). Lower-yield ports—such as port 80—are deprioritized or skipped entirely when unlikely to contribute. To quantify the impact of these tweaks, we evaluate four variants: T1 stops probing after the first valid certificate per host; T2 builds on T1 by prioritizing port 443; T3 further skips port 80 if port 443 does not yield a certificate; and T4 restricts scanning solely to port 443.

\myitem{Performance:}
As shown in Figure~\ref{fig:runtime-optimizations}, these tweaks significantly reduce runtime. On a single thread, T1 reduces worst-case scan time from 260 to under 39 minutes, while T4 brings it down to just 4 minutes. In terms of accuracy, the tweaks result in minimal witness loss (Figure~\ref{fig:optimizations-drops}). Across all prefixes, $<2$ witnesses were missed in 83.6\% of cases. The worst-case drop—13 witnesses in a high-density prefix—still left \sys with 15 usable witnesses.

These results show that \sys can operate in real-time in practice, and even coarse-grained probing (T4) is sufficient for building a meaningful witness set in most prefixes.
\section{Evaluation}\label{sec:evaluation}
\subsection*{Level of Protection}
\myitem{Methodology:} In this section, we evaluate \sys's ability to provide protection to prefixes of four privacy-sensitive applications. We define two levels of protection that \sys offers: \textit{Light} and \textit{Strong}. \sys offers Light protection if it can find at least 3 witnesses in the given prefix. To offer Strong protection, \sys needs to find a set of at least 8 witnesses that serve certificates from 2 or more different CAs between them.

\myitem{Bitcoin nodes:} Bitcoin node operators can employ \sys to improve the security of their connections. For example, an adversary capable of hijacking and impersonating the connections to a node can seek to eclipse it, enabling selfish mining and double spend attacks~\cite{heilman2015eclipse}. \sys can detect that the node's connections are being impersonated and alert the node operators to seek new peers.
We evaluate \sys's ability to protect a node's connections to its peers by sampling 200 random node IPs from a Bitcoin node tracker~\cite{bitnodes}. Our results, 
indicate that \sys can provide Light protection to 59\% of nodes, and Strong protection to 49\%.

\myitem{Bitcoin mining pool gateways:} Mining pool gateways are the Bitcoin nodes used by pools to propagate their freshly mined blocks. With each block offering a mining reward of over 337,000 USD at the time of writing, there is great incentive for pools to ensure that their blocks are quickly disseminated to other influential nodes in the network using low latency connections. An adversary seeking to damage a pool gateway's ability to broadcast blocks can hijack and impersonate its connections to other influential nodes. \sys can detect the impersonation attack and alert gateway operators to adjust their peers. We assess \sys's ability to protect connections to 44 nodes we inferred to be mining pool gateways using timing analysis. 
\sys extends Strong protection to 68\% of inferred gateway pool prefixes, and Light protection to 75\% of inferred gateway prefixes. 
We believe \sys extends better protection to pool gateways compared to random Bitcoin nodes because the former are more likely to be hosted on professional infrastructure and share a prefix with other services that can act as witnesses, rather than run by individual residential users.

\myitem{DNS Servers:} Despite efforts pushing for adoption, authentication in DNS is still widely not deployed~\cite{misell2025measuring}. An adversary can use \interception to redirect DNS clients to malicious servers. \sys can validate the destination prefix before each DNS request, to ensure that the response does not originate from an impersonating adversary. We evaluate \sys's ability to protect DNS clients by looking for witnesses in the prefixes of 200 sampled DNS servers~\cite{dns_resolvers}. We find that \sys can provide Light protection to 40\% of prefixes, and Strong protection to 12\% of prefixes.

\myitem{Popular Websites:} \interception attacks in which the adversary obtains a bogus certificate and serves a malicious copy of a website have been both studied in the literature~\cite{birge2018bamboozling}, and observed in the wild~\cite{myetherwallet_bgp}. \sys can alert users to such impersonation by observing witness failures. We evaluate \sys on protecting the top 200 websites on the Tranco list~\cite{LePochat2019}. \sys provides Light protection to 87\% of prefixes, and Strong to 55\%.

\myitem{Tor Relays:} To further understand \sys's ability to protect prefixes that host security and privacy sensitive applications that can be the target of routing attacks, we identify witnesses for a sample of 200 random relays. Our findings, illustrated in Figure~\ref{fig:applications}, show that \sys can offer Strong protection to 64\% of relay prefixes, and Light to 89\%.

\subsection*{Witness Reliability}
\myitem{Methodology:} We evaluate the reliability of \sys witnesses over time by finding and monitoring witnesses for the aforementioned applications over the course of 6 days. We identify 30217 witnesses spanning the 714 prefixes, which we poll in 15 minute intervals. Witnesses that fail 10 consecutive polls are discarded from the witness set. 

\myitem{Witness Latency:} \sys performs periodic witness polling in the background to maintain the witness set. However, there are applications (e.g web browsing) where a user might wish to also poll witnesses on demand. Figure~\ref{fig:violin} shows the latency distribution for all witnesses, and for witnesses according to their latency rank within their prefix. Given that witness polls can be parallelized, our results indicate including more witnesses in the witness set adds minimal overhead.

\myitem{Witness Uptime:} \sys depends on witnesses maintaining a high uptime. Figure~\ref{fig:witness_uptime_dist} illustrates the probability of a witness being online during poll time; 96\% of our final witness set was online for 99+\% of polls. Figure~\ref{fig:uptime_heatmap} shows the fraction of online witnesses for 40 sample prefixes over the span of 72 poll sequences. Witnesses . Finally, Figure~\ref{fig:offline_witnesses_ts} shows the number of witnesses from our witness set go offline over time. At the end of our 6 day monitoring period, just over 4\% of our witness set had gone offline, indicating that \sys does not require frequent witness re-discovery.

\begin{figure*}[t]
    \centering
    \begin{subfigure}[t]{0.45\linewidth}
        \centering
        \begin{tikzpicture}
        \begin{axis}[
            view={0}{90},
            xlabel={Poll Sequence},
            ylabel={Prefix Index},
            colorbar,
            colormap name=viridis,
            colorbar style={
                ytick={0, 0.2, 0.4, 0.6, 0.8, 1.0},
                yticklabel={
                    \pgfmathparse{100*\tick}
                    \pgfmathprintnumber[fixed, precision=0]{\pgfmathresult}\%
                }
            },
            point meta min=0,
            point meta max=1,
            width=\linewidth, 
            height=0.55\linewidth,
            enlargelimits=false,       
            yticklabels=\empty,        
            grid=none,
            axis on top,               
            tick label style={font=\small},
            mesh/cols=72,
        ]
        
        \addplot[
            matrix plot,
            point meta=\thisrow{online_fraction},
        ] table [x=poll_idx, y=prefix_idx, col sep=comma] {data/availability_uptime/prefix_uptime_heatmap.csv};
        
        \end{axis}
        \end{tikzpicture}
        \caption{Heatmap illustrating per-prefix witness availability over time.}
        \label{fig:uptime_heatmap}
    \end{subfigure}
    \hfill
    \begin{subfigure}[t]{0.45\linewidth}
        \centering
        \begin{tikzpicture}
        \begin{axis}[
            width=\linewidth, 
            height=0.55\linewidth,
            xlabel={Date},
            ylabel={Offline Witnesses},
            grid=major,
            xtick={0, 86400, 172800, 259200, 345600, 432000, 518400},
            xticklabels={0, 1, 2, 3, 4, 5, 6},
            scaled x ticks=false,
            xlabel={Day Number},
            xmin=0,
            xmax=550000,
            ymin=0,
            no marks,
            line join=round,
        ]
        
        \addplot[
            color=red,
            thick
        ] table [
            x expr=\thisrow{timestamp_sec}-1780179750.18, 
            y=cumulative_retired, 
            col sep=comma
        ]{data/availability_failures/retired_cumulative.csv};
        
        \end{axis}
        \end{tikzpicture}
        \caption{Witnesses that go offline over time.}
        \label{fig:offline_witnesses_ts}
    \end{subfigure}

    \vspace{1em}
    
    \begin{subfigure}{0.45\linewidth}
        \centering
        \begin{tikzpicture}
        \begin{axis}[
            ybar,
            width=\linewidth,
            height=0.55\linewidth,
            xlabel={Uptime (\%)},
            ylabel={Witnesses \# (Log)},
            xtick=data,
            table/col sep=comma,
            xticklabels from table={data/availability_uptime/uptime_binned.csv}{binlabel},
            ymode=log, 
            grid=major,
            ymin=1, 
            x tick label style={rotate=45, anchor=east, font=\scriptsize},
        ]
        \addplot[fill=blue!30] table [x expr=\coordindex, y=witnesscount, col sep=comma] {data/availability_uptime/uptime_binned.csv};
        \end{axis}
        \end{tikzpicture}
        
        \caption{96\% of witnesses have 99+\% uptime.}
        \label{fig:witness_uptime_dist}
    \end{subfigure}
    \hfill
    \begin{subfigure}{0.45\linewidth}
        \centering
        \begin{tikzpicture}
            \begin{axis}[
                ylabel={Latency (ms)},
                xlabel={Witness Rank},
                xtick={1,2,3,4,5,6},
                xticklabels={All, \#1, \#2, \#5, \#10, \#25},
                ymin=0,
                width=\linewidth,
                height=0.55\linewidth,
                grid=major,
                boxplot/draw direction=y,
            ]
            \newcommand{\violinwithbox}[9]{
                \addplot[#3!20, fill=#3!20, no markers, forget plot] 
                    table [y=latency_ms, x expr={#2 + \thisrow{rel_x}}, col sep=comma] 
                    {data/availability_latency/violin_boundary_#1.csv};
                \addplot[#3, thin, opacity=0.5, no markers, forget plot] 
                    table [y=latency_ms, x expr={#2 + \thisrow{rel_x}}, col sep=comma] 
                    {data/availability_latency/violin_boundary_#1.csv};
                \addplot [
                    black, thick,
                    boxplot prepared={
                        draw position=#2,
                        lower whisker=#4,
                        lower quartile=#5,
                        median=#6,
                        upper quartile=#7,
                        upper whisker=#8,
                        box extend=0.1,
                    },
                ] coordinates {};   
            }
            \violinwithbox{all}{1}{blue}{7.55}{44.49}{175.26}{211.08}{560.00}{182.54}
            \violinwithbox{k1}{2}{red}{7.55}{50.09}{184.45}{232.08}{503.04}{193.34}
            \violinwithbox{k2}{3}{orange}{9.42}{46.81}{184.42}{228.23}{542.02}{196.85}
            \violinwithbox{k5}{4}{green}{11.52}{45.33}{184.62}{231.79}{559.11}{202.76}
            \violinwithbox{k10}{5}{purple}{11.79}{42.29}{183.40}{210.39}{549.60}{187.69}
            \violinwithbox{k25}{6}{brown}{16.90}{41.74}{184.31}{205.70}{526.13}{184.91}
            \end{axis}
        \end{tikzpicture}
        \caption{Mean poll latency distribution by witness latency rank within witness prefixes.}
        \label{fig:violin}
    \end{subfigure}
\end{figure*}
\section{Limitations}

\sys’s effectiveness depends on the presence of authenticatable, witnesses in the destination prefix. While our findings (\S\ref{sec:measurements}, \S\ref{sec:evaluation}) show this is often the case, it is not universal.
\sys also assumes a benign initial state when establishing the witness set. If initialized during an attack, adversarial witnesses may be incorporated. While an attacker could attempt to introduce malicious witnesses within the destination prefix, doing so requires control over infrastructure in that prefix, or the cloud provider that presides over that prefix allowing customers to hand-pick prefixes, which is not always feasible.
Finally, \sys does not defend against routing adversaries capable of forwarding traffic (e.g.,~\cite{birge2019sico}), which can preserve consistency across witnesses. Such attacks are more complex in their requirements and may be mitigated using additional signals such as latency or observed path anomalies.
\section{Future Directions}
\label{sec:future}

\myitem{Online security through \sys-native browsers:}
When a browser visits a website and verifies its TLS certificate, it could trigger \sys to discover witnesses within the same IP prefix to detect potential BGP hijacks. However, browsers currently lack a notion of IP prefixes and would require mechanisms to infer or query them. Additionally, probing other services in the prefix may introduce privacy concerns, increase page load latency, or trigger firewall and rate-limiting defenses. Finally, the browser would need to maintain a distributed database of verifiable services per prefix. 
\emph{Research Question: What mechanisms can enable prefix-wide validation with minimal impact on latency and resource usage?}

\myitem{\sys-Mesh: checking reachability from the Internet's edge:}
Rather than relying on individual hosts to discover and verify witnesses, a set of distributed clients (trust anchors) could periodically verify reachability and authenticity among themselves. Collectively, they could build a public map of validated Internet connectivity, similar to PingMesh in datacenters. Such a system could leverage proof-of-stake protocols, where participants attest to verified reachability and are incentivized accordingly. This may also address the scarcity of stable witnesses by encouraging hosts without public services to participate by exposing verifiable identities. 
\emph{Research Question: How can we design a decentralized system that continuously validates Internet reachability and authenticity across distributed vantage points and exposes disruptions (e.g., BGP hijacks) in real time?}

\myitem{Identifying witnesses beyond TLS certificates:}
While this work focuses on websites due to the ease of TLS verification, other observable traits can serve as fingerprints, including open ports, hardware characteristics, or latency profiles. This enables multi-modal verification strategies that may be client-specific and raise the bar for adversaries attempting to impersonate a prefix.
\emph{Research Question: Can combining such signals make destinations harder to convincingly fake during a hijack?}
\section*{Acknowledgments}
We thank Censys~\cite{censys15} for granting us access to their TLS certificate history data, which was used to verify the effectiveness of \sys.

\printbibliography

\section*{Ethics}
\label{sec:ethics}
This work involves measurements of public-facing hosts on the internet, and proposes a system for detecting routing-based impersonation attacks. We consider the ethical implications of both our data collection and the potential deployment of \sys.

Our analysis relied on commercially accessible datasets (e.g., Censys) and data collected from our own isolated scans. The data we use consists of service metadata (e.g., TLS certificates) and does not include user-generated content or personally identifiable information.

\sys performs lightweight reachability and authentication checks to services within a prefix. These interactions resemble standard client behavior (e.g., TLS handshakes) and are designed to minimize overhead. To minimize the potential effects our measurements may have had to remote hosts, we implemented strict rate limiting.

While \sys is designed to detect impersonation attacks, similar techniques could be used to map service deployments within prefixes or probe infrastructure characteristics. We do not release any tools that enable large-scale probing beyond what is already possible with existing scanning platforms, and we advocate for responsible use of such techniques.

Our study does not identify or exploit specific vulnerable systems.
\section*{Appendix}
\myitem{Witness Poll Failures:} More than 55\% of witness polling failures are attributed to timeout errors possibly stemming either from the witness being unavailable, or intervention by middleboxes. We also frequently encountered instances where the connection to the witness was either interrupted during the TLS handshake (34\%) or directly refused (3.8\%). For a summary of the most frequent failures see Table~\ref{tab:witness_errors}.
\begin{table}[]
    \centering
    \begin{adjustbox}{max width=0.5\linewidth}
        \begin{filecontents*}{data/availability_failures/failure_modes.csv}
layer,errortype,count,percentage
TCP,TCP Dial Timeout,154991,38.26
TLS,Connection Reset (Read),138872,34.28
TCP,Read Timeout,69077,17.05
TCP,Connection Refused,15401,3.8
TCP,No Route to Host,9545,2.36
TLS,Unexpected EOF,6039,1.49
TLS,TLS: Internal Error,4691,1.16
TLS,TLS: Handshake Failure,2458,0.61
TLS,TLS: Unrecognized Name,2377,0.59
TLS,Not a TLS Handshake,1174,0.29
TCP/TLS,Other,473,0.12
\end{filecontents*}
\pgfplotstabletypeset[
            col sep=comma,
            string type,
            columns/layer/.style={
                column name=\textbf{Layer}, 
                column type=l
            },
            columns/errortype/.style={
                column name=\textbf{Error Description}, 
                column type=l
            },
            columns/count/.style={
                column name=\textbf{Count}, 
                column type=r,
                string type,
            },
            columns/percentage/.style={
                column name=\textbf{\%},
                column type=r
            },
            every head row/.style={
                before row=\toprule, 
                after row=\midrule
            },
            every last row/.style={
                after row=\bottomrule
            },
        ]{data/availability_failures/failure_modes.csv}
    \end{adjustbox}
    \caption{Distribution of witness poll failure errors.}
    \label{tab:witness_errors}
\end{table}

\end{document}